# The effect of bound states on X-ray Thomson scattering for partially ionized plasmas

J. Nilsen, W. R. Johnson, K. T. Cheng

# The effect of bound states on X-ray Thomson scattering for partially ionized plasmas

Joseph Nilsen[1], Walter R. Johnson[2], and K. T. Cheng[1]

[1]Lawrence Livermore National Laboratory, Livermore, CA 94551
[2]University of Notre Dame, Notre Dame, IN 46556

**Abstract.** X-ray Thomson scattering is being developed as a method to measure the temperature, electron density, and ionization state of high energy density plasmas such as those used in inertial confinement fusion. X-ray laser sources have always been of interest because of the need to have a bright monochromatic x-ray source to overcome plasma emission and eliminate other lines in the background that complicate the analysis. With the advent of the x-ray free electron laser (X-FEL) at the SLAC Linac Coherent Light Source (LCLS) and other facilities coming online worldwide, we now have such a source available in the keV regime. An important challenge with x-ray Thomson scattering experiments is understanding how to model the scattering for partially ionized plasmas. Most Thomson scattering codes used to model experimental data greatly simplify or neglect the contributions of the bound electrons to the scattered intensity. In this work we take the existing models of Thomson scattering that include elastic ion-ion scattering and inelastic electron-electron scattering and add the contribution of bound electrons in the partially ionized plasmas. Except for hydrogen plasmas, most plasmas studied today have bound electrons and it is important to understand their contribution to the Thomson scattering, especially as new x-ray sources such as the X-FEL will allow us to study much higher Z plasmas. To date, most experiments have studied hydrogen or beryllium plasmas. We first analyze existing experimental data for beryllium to validate the code. We then consider several higher Z materials such as Cr and predict the existence of additional peaks in the scattering spectrum that requires new computational tools to understand. For a Sn plasma, we show that bound contributions change the shape of the scattered spectrum in a way that would change the plasma temperature and density inferred from experiment.

## 1. Introduction

X-ray Thomson scattering is being used as an important diagnostic technique to measure temperatures, densities, and ionization balance in warm dense plasmas. Glenzer and Redmer [1] have reviewed the underlying theory of Thomson scattering used to analyze experiments.

We start with the theoretical model proposed by Gregori et al. [2] but evaluate the Thomson-scattering dynamic structure function using parameters taken from our own average-atom code [3,4]. The average-atom model is a quantum mechanical version of the temperature-dependent Thomas-Fermi model of plasma developed years ago by Feynman et al. [5]. It consists of a single ion of charge Z with a total of Z bound and continuum electrons in a Wigner-Seitz cell that is embedded in a uniform "jellium sea" of free electrons whose charge is balanced by a uniform positive background. This model enables us to consider the contributions from the bound electrons in a self-consistent way for any ion. Other approaches such as Gregori's employ hydrogenic wave functions with screening factors to approximate the contribution from the bound electrons for a limited number of materials.

In this paper, we analyze existing experimental data [6] for beryllium and use this to validate our code. We then consider several higher Z materials such as Cr and predict the existence of additional peaks in the scattering spectrum that requires new computational tools to understand. For a Sn plasma, we show that contributions from the bound electrons can change the shape of the scattered spectrum in a way that would change the plasma temperature and density inferred by the experiment.

**2. Overview of theory**

A detailed description of the theory used in this paper can be found in Ref. [4]. Briefly, the Thomson scattering cross section is proportional to the dynamic structure function $S(k,\omega)$, where $\hbar k$ and $\hbar\omega$ are the momentum and energy transfers, respectively, from the incident to the scattered photons. (For simplicity, $\hbar$ will be dropped in the following.) As shown in the work of Chihara [7,8], $S(k,\omega)$ can be decomposed into three terms: the first term $S_{ii}(k,\omega)$ is the contribution from elastic scattering by electrons that follow the ion motion, the second term $S_{ee}(k,\omega)$ is the contribution from inelastic scattering by free electrons, and the third term $S_b(k,\omega)$ is the contribution from bound-free transitions (inelastic scattering by bound electrons) modulated by the ionic motion. In the present work, the modulation factor is ignored when evaluating the bound-free scattering structure function. For the bound-free contribution, our calculations use average-atom scattering wave functions for the final states. In Ref. [4] we show that using plane-wave final states can give very different results that disagree with experimental data.

The theoretical model developed by Gregori et al. [2] is used to evaluate the ion-ion contribution $S_{ii}(k,\omega)$ to the dynamic structure function, with scattering form factors (Fourier transforms of the charge densities) calculated with bound- and continuum-state wave functions from our average-atom code. Additionally, the procedure proposed in Ref. [9] is used to account for differences between electron and ion temperatures. The electron-electron contribution $S_{ee}(k,\omega)$ is expressed in terms of the dielectric function $\varepsilon(k,\omega)$ of the free electrons, which in turn is evaluated using the random-phase approximation (RPA) as in Ref. [2]. Finally, bound-state contributions to the dynamic structure function are evaluated using average-atom bound- and continuum-state wave functions, with the latter approaching plane waves asymptotically.

In the collective regime, which is for small momentum transfers and usually for forward angle scattering, one sees plasmon peaks that are up and down shifted in energy from the central peak. For experiments that can observe both peaks, the plasma temperature can be determined from the ratio of the two plasmon peaks by exp[-ΔE/kT] where ΔE is the energy shift of the plasmon peak [1,4]. The energy shift of the plasmon peak energy has a plasma frequency component that is proportional to the square root of the free-electron density but it also has a thermal energy contribution and Compton shift that depend on the x-ray energy, temperature, and scattering angle [1]. As a result, the electron density is usually determined by doing a best fit to the experimental data.

## 3. Validating the average-atom code for Be experiments

Numerous experiments have been done to look at Thomson scattering for low-Z materials such as hydrogen and beryllium. We examine one particular experiment [6] done at the Omega laser facility that looked at Thomson scattering in the forward scattering direction (40 degrees) from solid Be with a Cl Ly-α source at 2963 eV. Figure 1 shows the measured spectrum as the noisy dotted line. An electron temperature of 18 eV and density of 1.647 g/cc is used in the average-atom model to give an electron density of 1.8 x$10^{23}$ per cc, in agreement with the analysis in Ref. [6]. The dashed line shows the experimental source function from the Cl Ly-α line. Because of satellite structure we approximate the source by 3 lines: a Cl Ly-α line at 2963 eV with amplitude 1.0 and two satellites at 2934 and 2946 eV with relative amplitudes of 0.075 and 0.037 respectively. Using the 3 weighted lines to do the Thomson scattering calculation, we calculate the scattering amplitude for Thomson scattering (solid line) and compare against the experimental data (dotted line). We observe excellent agreement within the experimental noise. Contributions from the bound 1s electrons have a threshold at 2875 eV that is beyond the range of the data shown in Fig. 1. A low ion temperature of 2.1 eV was needed in order to match the experimental data for the central peak.

For this Be plasma, the plasmon energy is calculated to be 26.95 eV. There are three components [1] summed in quadrature to get this value. The plasma frequency due to the free electrons is 15.7 eV, the thermal energy contribution is 21.5 eV, and the Compton shift is 4.0 eV. The Fermi energy at the Be density is 11.6 eV which means we are in a regime where the temperature is greater than the Fermi energy and can use the coherence parameter $\alpha = 1/(k\lambda_D)$ to determine if we are in the collective regime where $\alpha > 1$. For the 40 degree angle used in experiment, the momentum transfer k = 1.026 Å$^{-1}$ while the Debye length $\lambda_D$ is 0.744 Å results in $\alpha$ = 1.3. If we looked at backward scattering angle of 130 degree, $\alpha$ drops to 0.5 and we would be in the non-collective regime and expect the plasmon peaks to disappear.

For Be under the conditions just described, the L-shell is completely stripped but the K-shell is 97% occupied. The average-atom code has $Z_f$ = 1.647 for the number of free electrons per ion in the jellium sea outside the Wigner-Seitz cell. Inside the Wigner-Seitz cell we have only 2 bound 1s electrons and 2 continuum electrons. The difference between the 2 continuum electrons and the 1.647 free electrons per ion outside the cell is due to the fact that the continuum charge density integrates to a charge of 2 *inside* the Wigner-Seitz cell but continues smoothly into the uniform free-electron density *outside* the cell with $Z_f$ = 1.647. A more detailed explanation can be found in Ref. [4]. Our model is using the free electrons per ion $Z_f$ for the $S_{ee}$ term so this is an important issue to understand. The binding energy of the K-shell electron is 87.6 eV which means the threshold for seeing the bound state contribution is at 2963 – 87.6 = 2875.4 eV. To understand the contributions of the bound K-shell electrons to the Thomson scattering, we expand the energy scale in Fig. 1 and re-plot it on a log-scale in Fig. 2, looking just at the average-atom calculation. We observe a K-shell contribution that is about 40 times weaker than the low-energy plasmon peak so it would be very difficult to observe in a laser-plasma experiment given the experimental noise.

## 4. Modelling Cr and Sn experiments

If we now consider higher-Z materials such as Cr and Sn at an electron temperature of 10 eV, we find that the bound state contribution can be very important. Following Refs. [4,6] and our analysis of the Be experiment, an ion/electron temperature ratio of 0.1 is assumed to reduce the central elastic scattering peak. The average-atom code predicts that solid density Cr (7.19 g/cc) heated to 10 eV has 6.2 continuum electrons in the Wigner-Seitz cell. This makes it a closed Ar-like core consisting of nearly fully occupied 3s and 3p subshells with binding energies of 57.7 and 30.7 eV, respectively. The code also predicts $Z_f$ = 2.92 and an electron density of 2.4 x 10$^{23}$ per cc. This is a case where there is a factor of 2 difference between the number of continuum electrons in the Wigner-Seitz cell and the asymptotic $Z_f$ value. For codes such as Gregori's the number of free electrons is an input parameter so it is important to understand what is the best value to choose.

Figure 3 shows the scattered intensity versus photon energy for calculations done with (solid line) and without (dashed line) the contribution of bound electrons for a 4750 eV x-ray source scattered at 40° off Cr. Without bound electrons, we predict the central elastic scattering peak from $S_{ii}(k,\omega)$ and the plasmon peaks from $S_{ee}(k,\omega)$. When we include the effect of bound electrons, we predict a very strong scattering peak that is downshifted by about 40 eV from the central elastic peak due to the 3p electrons and a much weaker 3s peak at lower energy.

For solid density Sn (7.3 g/cc) at 10 eV we look at the case of the 2960 eV x-ray source scattering in the backward direction at 130°. The average-atom code predicts that warm dense Sn has 4.4 continuum electrons with an average occupation of 8.8 for the 4d electrons and 0.8 for the 5s electrons outside a closed Kr-like core with all the lower orbital subshells fully occupied. We predict $Z_f = 3.37$ which gives an electron density of $1.25 \times 10^{23}$ per cc. The binding energies of the 4d and 5s electrons are 22.4 and 2.14 eV, respectively, which means the effect of the 5s electron will be hidden under the elastic scattering peak. Fig. 4 plots the scattering intensity versus photon energy for calculations done with (solid line) and without (dashed line) the contribution of the bound electrons. For backward scattering the scattering is no longer in the collective regime and the distinct plasmon peaks are replaced by a broader scattering structure. The contribution from the bound 4d electrons is now an additional large broad feature that lies on top of the Thomson scattering and would make it easy to misinterpret the spectrum if one did not understand the contribution of the bound electrons.

## 5. Conclusions

The availability of bright monochromatic x-ray line sources from x-ray free electron laser facilities opens up many new possibilities to use Thomson scattering as an important diagnostics technique to measure the temperatures, densities, and ionization balance in warm dense plasmas.

Current attempts to model Thomson scattering tend to use very simplified models, if anything, to model the effect of the bound electrons on the measured scattered intensity. Our approach here is to evaluate the Thomson-scattering dynamic structure function using parameters taken from our own average-atom code [3,4]. This model enables us to consider contributions from the bound electrons in a self-consistent way for any ion.

In this paper we validate our average-atom based Thomson scattering code by comparing our model against existing experimental data for Be near solid density and temperature near 18 eV. For solid density Cr at 10 eV we predict the existence of additional peaks in the scattering spectrum that requires new computational tools to understand. We also analyse solid density Sn at 10 eV and show that the contributions from the bound electrons can change the shape of the scattered spectrum in a way that would change the plasma temperature and density inferred by the experiment.

**Acknowledgements.** This work was performed under the auspices of the U.S. Department of Energy by Lawrence Livermore National Laboratory under Contract DE-AC52-07NA27344.

**Figure captions:**

**Fig. 1.** (Color online) Intensity vs scattered photon energy for scattering of a Cl Ly-α x-ray from Be at 40°. Dotted line is experimental data (see Ref. [6]) while solid line is calculation for an electron temperature of 18 eV. The dashed line shows the Cl x-ray source.

**Fig. 2.** Intensity vs scattered photon energy for calculation of scattering of a Cl Ly-α x-ray from Be at 40° for an electron temperature of 18 eV. The contribution from the K-shell bound electrons is shown.

**Fig. 3.** (Color online) Intensity vs scattered photon energy for calculation of scattering of a 4750 eV x-ray line from Cr at 40° for an electron temperature of 10 eV. The case where the 3p and 3s bound electron contributions are included is shown by solid line and case without bound electrons is shown by dashed line.

**Fig. 4.** (Color online) Intensity vs scattered photon energy for calculation of scattering of a 2960 eV x-ray line from Sn at 130° for an electron temperature of 10 eV. The strong contribution from the 4d bound electrons are apparent for the case where the bound electrons are included (solid line) when compared with case without bound electrons (dashed line).

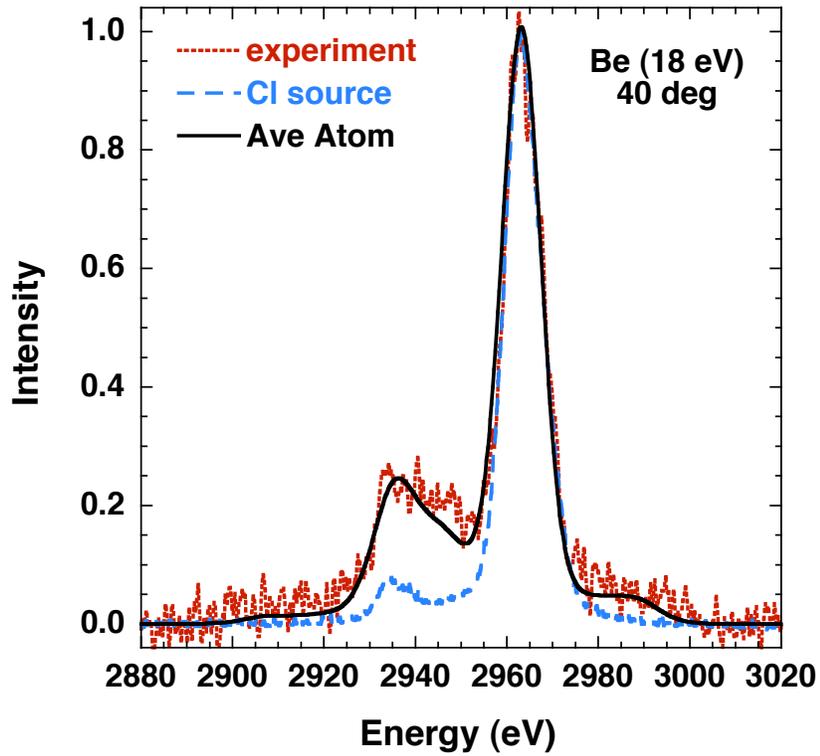

**Fig. 1.** (Color online) Intensity vs scattered photon energy for scattering of a Cl Ly-α x-ray from Be at 40°. Dotted line is experimental data (see Ref. [6]) while solid line is calculation for an electron temperature of 18 eV. The dashed line shows the Cl x-ray source.

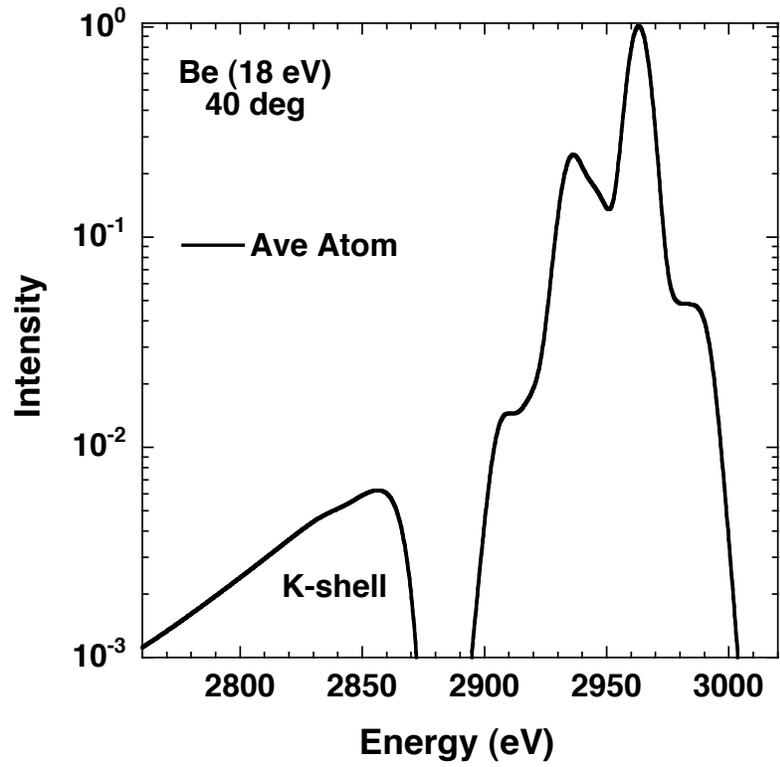

**Fig. 2.** Intensity vs scattered photon energy for calculation of scattering of a Cl Ly-α x-ray from Be at 40° for an electron temperature of 18 eV. The contribution from the K-shell bound electrons is shown.

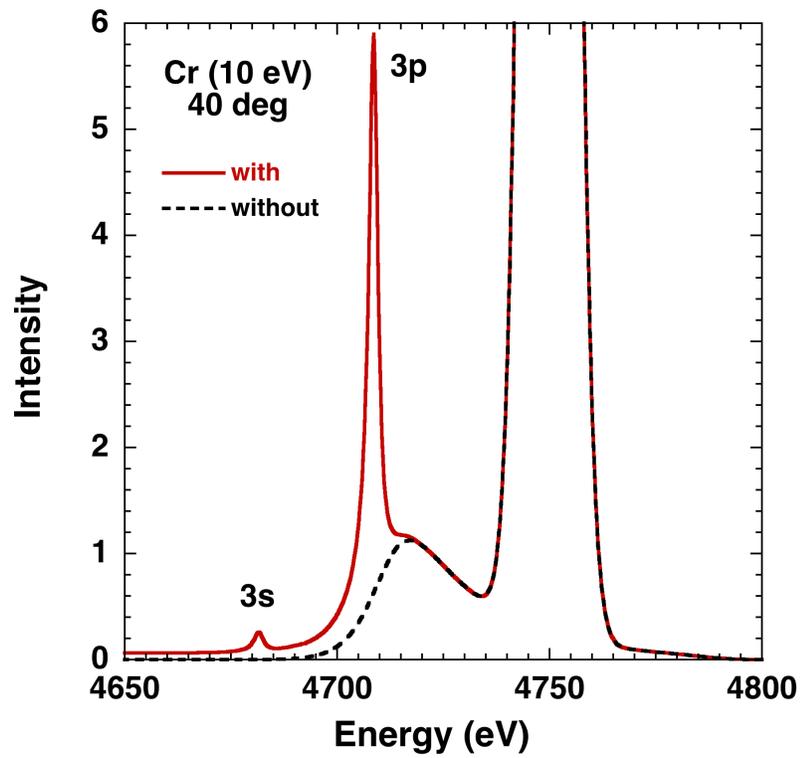

**Fig. 3.** (Color online) Intensity vs scattered photon energy for calculation of scattering of a 4750 eV x-ray line from Cr at 40° for an electron temperature of 10 eV. The case where the 3p and 3s bound electron contributions are included is shown by solid line and case without bound electrons is shown by dashed line.

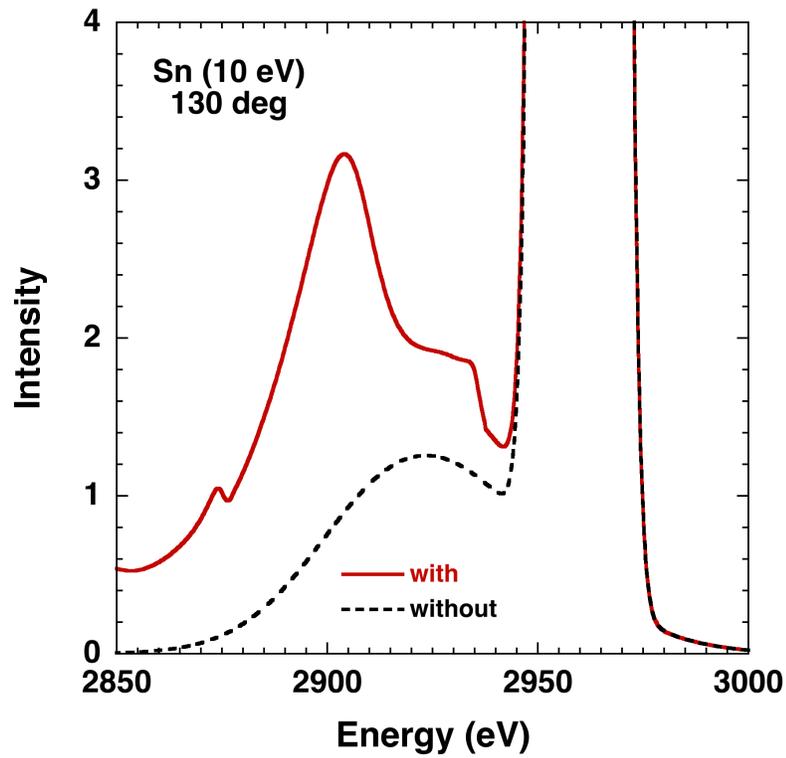

**Fig. 4.** (Color online) Intensity vs scattered photon energy for calculation of scattering of a 2960 eV x-ray line from Sn at 130° for an electron temperature of 10 eV. The strong contribution from the 4d bound electrons are apparent for the case where the bound electrons are included (solid line) when compared with case without bound electrons (dashed line).